\documentclass[preprint,aps,showpacs,nofootinbib,prd,superscriptaddress]{revtex4-1}

\usepackage{graphicx}
\usepackage{amsfonts}
\usepackage[mathscr]{eucal}
\usepackage{epsfig,amssymb,amsmath} 
\usepackage{epstopdf}
\usepackage{times}
\usepackage{dcolumn}
\usepackage{bm}
\usepackage[colorlinks=true,citecolor=blue,linkcolor=blue]{hyperref}

\usepackage{hyperref}
\usepackage{color}
\usepackage{cancel}

\begin{document}


\newcommand{\vev}[1]{ \left\langle {#1} \right\rangle }
\newcommand{\bra}[1]{ \langle {#1} | }
\newcommand{\ket}[1]{ | {#1} \rangle }
\newcommand{\eV}{ \ {\rm eV} }
\newcommand{\KeV}{ \ {\rm keV} }
\newcommand{\MeV}{\  {\rm MeV} }
\newcommand{\GeV}{\  {\rm GeV} }
\newcommand{\TeV}{\  {\rm TeV} }
\newcommand{\1}{\mbox{1}\hspace{-0.25em}\mbox{l}}
\newcommand{\Red}[1]{{\color{red} {#1}}}

\newcommand{\lmk}{\left(}
\newcommand{\rmk}{\right)}
\newcommand{\lkk}{\left[}
\newcommand{\rkk}{\right]}
\newcommand{\lhk}{\left \{ }
\newcommand{\rhk}{\right \} }
\newcommand{\del}{\partial}
\newcommand{\la}{\left\langle}
\newcommand{\ra}{\right\rangle}
\newcommand{\half}{\frac{1}{2}}

\newcommand{\bea}{\begin{array}}
\newcommand{\eea}{\end{array}}
\newcommand{\beq}{\begin{eqnarray}}
\newcommand{\eeq}{\end{eqnarray}}

\newcommand{\dd}{\mathrm{d}}
\newcommand{\Mpl}{M_{\rm Pl}}
\newcommand{\mg}{m_{3/2}}
\newcommand{\abs}[1]{\left\vert {#1} \right\vert}
\newcommand{\mphi}{m_{\phi}}
\newcommand{\Hz}{\ {\rm Hz}}
\newcommand{\for}{\quad \text{for }}
\newcommand{\Min}{\text{Min}}
\newcommand{\Max}{\text{Max}}
\newcommand{\Kahler}{K\"{a}hler }
\newcommand{\cphi}{\varphi}
\newcommand{\Tr}{\text{Tr}}
\newcommand{\diag}{{\rm diag}}

\newcommand{\SUf}{SU(3)_{\rm f}}
\newcommand{\Upq}{\rm U(1)_{PQ}}
\newcommand{\Zpq}{Z^{\rm PQ}_3}
\newcommand{\Cpq}{C_{\rm PQ}}
\newcommand{\ubar}{u^c}
\newcommand{\dbar}{d^c}
\newcommand{\ebar}{e^c}
\newcommand{\nubar}{\nu^c}
\newcommand{\Ndw}{N_{\rm DW}}
\newcommand{\Fpq}{F_{\rm PQ}}
\newcommand{\fpq}{v_{\rm PQ}}
\newcommand{\Br}{{\rm Br}}
\newcommand{\Lag}{\mathcal{L}}
\newcommand{\Lqcd}{\Lambda_{\rm QCD}}

\newcommand{\ji}{j_{\rm inf}}
\newcommand{\jb}{j_{B-L}}
\newcommand{\M}{M}
\newcommand{\im}{{\rm Im} }
\newcommand{\re}{{\rm Re} }

\def\lrf#1#2{ \left(\frac{#1}{#2}\right)}
\def\lrfp#1#2#3{ \left(\frac{#1}{#2} \right)^{#3}}
\def\lrp#1#2{\left( #1 \right)^{#2}}
\def\REF#1{Ref.~\cite{#1}}
\def\SEC#1{Sec.~\ref{#1}}
\def\FIG#1{Fig.~\ref{#1}}
\def\EQ#1{Eq.~(\ref{#1})}
\def\EQS#1{Eqs.~(\ref{#1})}
\def\TEV#1{10^{#1}{\rm\,TeV}}
\def\GEV#1{10^{#1}{\rm\,GeV}}
\def\MEV#1{10^{#1}{\rm\,MeV}}
\def\KEV#1{10^{#1}{\rm\,keV}}
\def\blue#1{\textcolor{blue}{#1}}
\def\red#1{\textcolor{blue}{#1}}

\newcommand{\fa}{f_{a}}
\newcommand{\osc}{_{\rm osc}}
\newcommand{\bear}{\begin{array}}
\newcommand {\eear}{\end{array}}

\newcommand{\mav}{\left. m_a^2 \right\vert_{T=0}}
\newcommand{\mat}{m_{a, {\rm QCD}}^2 (T)}
\newcommand{\mam}{m_{a, {\rm M}}^2 }
\newcommand{\ds}{\displaystyle}
\newcommand{\non}{\nonumber}


\preprint{
TU-1017,\, \\
IPMU16-0030,\, \\
APCTP Pre 2016-006,\, \\
PNUTP-16/A11
}
\title{
Quality of the Peccei-Quinn symmetry 
in the Aligned QCD Axion \\
and  Cosmological Implications
}
\author{
 Tetsutaro Higaki
}
\affiliation{Department of Physics, Keio University, Kanagawa 223-8522, Japan}
\author{
Kwang Sik Jeong
}
\affiliation{Department of Physics, Pusan National University, Busan 46241, Korea
}
\author{
Naoya Kitajima
}
\affiliation{Asia Pacific Center for Theoretical Physics, Pohang 790-784, Korea}
\author{
Fuminobu Takahashi
}
\affiliation{Department of Physics, Tohoku University,
Sendai, Miyagi 980-8578, Japan}
\affiliation{Kavli IPMU (WPI), UTIAS,
The University of Tokyo,
Kashiwa, Chiba 277-8583, Japan}


\begin{abstract}
We show that the required high quality of the Peccei-Quinn symmetry can be naturally
explained in the aligned QCD axion models where the QCD axion arises from multiple axions with
decay constants much smaller than the axion window, e.g., around the weak scale.
Even in the presence of general Planck-suppressed Peccei-Quinn symmetry breaking operators,
the effective strong CP phase remains sufficiently small in contrast to the standard
axion models without the alignment.
The QCD axion potential has small or large modulations due to the symmetry breaking operators, 
which can significantly affect the axion cosmology.  
When the axions are trapped in different minima, domain walls appear and their scaling behavior
suppresses the axion isocurvature perturbations at super-horizon scales.  
Our scenario predicts many axions and saxions coupled to gluons, and they may
be searched for at collider experiments. In particular, the recently found diphoton excess at $750$~GeV
could be due to one of such (s)axions.
\end{abstract}

\maketitle

\section{Introduction}
The strong CP phase $\bar \theta$ is tightly constrained by the search for neutron electric dipole 
moment (EDM)~\cite{Baker:2006ts},
\beq
\label{theta}
|\bar \theta| \lesssim 10^{-10}.
\eeq
Why $\bar \theta$ is so small is known as the strong CP problem.
The strong CP problem is one of the remaining mysteries of the Standard Model (SM), and one plausible solution
is the Peccei-Quinn (PQ) mechanism~\cite{Peccei:1977hh,Weinberg:1977ma}. 
In association with spontaneous breakdown of a global PQ symmetry,
a pseudo Nambu-Goldstone (NG) boson, the QCD axion, appears.  If the PQ 
symmetry is explicitly 
broken only by the QCD instanton effects,  the QCD axion is stabilized at 
a CP conserving minimum, solving the strong CP problem. 
The QCD axion in the form of coherent oscillations is necessarily produced 
by the dynamical cancellation of the strong CP phase,
and it can account for the observed dark matter (DM) abundance. 

While elegantly solving the strong CP problem, the PQ mechanism poses two potential problems. 
One is the origin of the (classical) axion window at an intermediate scale,
\beq
\GEV{9} \lesssim F_a \lesssim \GEV{12},
\label{AW}
\eeq
where the lower bound is due to the observation of the SN 1987A neutrino burst duration~\cite{Mayle:1987as,Raffelt:1987yt,Turner:1987by},
and the upper bound is due to the axion contribution to the DM abundance barring fine-tuning of
the initial misalignment~\cite{Preskill:1982cy,Abbott:1982af,Dine:1982ah}. 
The origin of the PQ scale at an intermediate scale remains unknown. 
It may arise from some combinations of the supersymmetry  (SUSY) breaking scale and the Planck scale~\cite{Murayama:1992dj,Choi:1996vz,Asaka:1998ns,Jeong:2011xu}. 
On the other hand, there appear many moduli and axion fields
in string theory at the compactification of extra dimensions, and one of them
may be identified with the QCD axion. In this case the natural scale for the axion decay constant  is of order the
string scale, $F_a \sim 10^{15-16}$~GeV, if the compactification scale is comparable to the Planck scale.
Such a large axion decay constant generically leads to overproduction of
the axion DM.\footnote{The axion abundance can be suppressed by 
the anthropic selection of the initial misalignment~\cite{Tegmark:2005dy}, late-time entropy dilution~\cite{Steinhardt:1983ia,Kawasaki:1995vt,Kawasaki:2004rx}, or  early oscillations and adiabatic suppression due to extra PQ breaking terms~\cite{Takahashi:2015waa,Kawasaki:2015lpf}.}

The other problem is the required high quality of the PQ symmetry~\cite{Carpenter:2009zs}. 
In general,  a global symmetry is
considered to be explicitly broken in the quantum gravity theory~\cite{global-symmetry-gravity}, and so, 
we naively expect that there are
various PQ breaking operators suppressed by powers of the Planck mass. 
However, such extra PQ breaking terms tend to give too large contributions to the strong CP phase, 
spoiling the PQ mechanism. One can suppress dangerous
operators by imposing discrete symmetry $Z_N$ with large $N$, but the existence of such large discrete symmetry
may be implausible.
Thus, the required high quality of PQ symmetry is a puzzle in the low-energy four dimensional 
theory.\footnote{
String theory may provide a theoretical framework to address 
this question~\cite{Barr:1985hk,Choi:2011xt,Choi:2014uaa}.
}
One interesting possibility is that such high quality of the PQ symmetry is due to the requirement that the axion should
explain the present DM abundance~\cite{Carpenter:2009zs}, and we will return to this issue later in this paper.

The axion decay constant is not necessarily in one-to-one correspondence with the associated PQ breaking scale.
The effective axion decay constant for multiple PQ scalars with arbitrary PQ charges was studied in Ref.~\cite{Sikivie:1986gq},
where it was shown that the effective axion decay constant sensitively depends on the PQ charge assignment.
It was pointed out in Ref.~\cite{Kim:2004rp} that the axion decay constant can be 
enhanced by a factor of the largest hierarchy among the PQ charges  in a model with two axions and it 
was used to implement natural inflation with a super-Planckian decay constant. 
The enhancement is due to the alignment of the axion potentials.
The alignment mechanism with multiple axions was first studied in Ref.~\cite{Choi:2014rja}, where they showed
that an exponentially large 
enhancement is possible without introducing extremely large coefficients of the axions. This is because multiple axions with a certain combination split the required large PQ charges into many U(1) charges with a moderate size.
The alignment mechanism with multiple axions and various number
of symmetry breaking terms was studied subsequently in Refs.~\cite{Higaki:2014pja,Higaki:2014mwa},
where many axions form the axion landscape (see also Refs.~\cite{Wang:2015rel,Masoumi:2016eqo}). 
The linear realization of the alignment mechanism with two axions was first studied in Ref.~\cite{Harigaya:2014eta} and later extended to
multiple fields~\cite{Harigaya:2014rga}, where a peculiar structure of the U(1) charge assignment was noted. A more concrete realization
along this line was given in Refs.~\cite{Choi:2015fiu,Kaplan:2015fuy,Fonseca:2016eoo}, and it was coined a clockwork axion model.

We have recently proposed a QCD axion model based on the alignment mechanism with clockwork structure~\cite{Higaki:2015jag},
where one of the axions or saxions can account for the recently found $750$~GeV diphoton excess~\cite{ATLAS,CMS}. 
In Ref.~\cite{Higaki:2015jag} we briefly discussed the quality of the PQ 
symmetry. One of the striking features of the
aligned QCD axion model is that the actual symmetry breaking scale can be much smaller than the conventional axion window (\ref{AW}).
As a result, any Planck-suppressed PQ breaking operators are highly suppressed compared to the usual scenario.\footnote{ 
In the models considered in Refs.~\cite{Rubakov:1997vp,Berezhiani:2000gh,Hook:2014cda,Fukuda:2015ana,Albaid:2015axa,Chiang:2016eav},
where the PQ breaking scale is low and corresponds to the axion decay constant, 
the QCD axion is visible but has a heavy mass to avoid the astrophysical constraints.
}
The high quality of PQ symmetry is a natural outcome of the aligned QCD axion model.

In this paper we study in detail both phenomenological and cosmological implications of
the aligned QCD axion model, which is based on an effective field theory approach.
The purpose of this paper is twofold. First we study the effect of  Planck-suppressed 
PQ breaking operators in detail in the aligned QCD axion model. In particular, we find a new regime where the axion mass mainly arises from
PQ breaking terms, while the strong CP phase remains sufficiently small. Interestingly, the QCD axion can have a mass much heavier
than in the conventional scenario. Secondly, the axion cosmology can be significantly modified
by such PQ breaking operators, which induce small or large modulations on the axion potential.
We investigate the axion cosmology such as the axion DM and its isocurvature perturbations in the presence of 
PQ breaking operators. In extreme cases, the QCD axion can be cosmologically unstable, decaying into hidden photons.

The rest of this paper is organized as follows.
We review the QCD axion model based on the alignment mechanism in Section~\ref{sec:2}, and then
discuss in Section~\ref{sec:3} how it helps to explain the high quality of PQ symmetry at low energy scales. 
We explore the QCD axion dynamics in the early Universe in Section~\ref{sec:4}.  
The contents of Sections \ref{sec:3} and \ref{sec:4} are our main new results.
Section~\ref{sec:5} is devoted to discussion and conclusions.

\section{Aligned QCD axion}
\label{sec:2}

In this section we first review the aligned axion model~\cite{Kim:2004rp,Choi:2014rja,
Higaki:2014pja,Harigaya:2014rga,Choi:2015fiu,Kaplan:2015fuy}, and apply the idea to the QCD
axion to see how the QCD axion could arise from multiple axions with low axion decay constants 
through the alignment mechanism. 


The alignment mechanism~\cite{Choi:2014rja} is implemented by multiple periodic axions,  
\beq
\phi_i \equiv \phi_i + 2\pi f_i  \quad(i=1,2,...,N)
\eeq 
with the potential of the form
\beq
\label{alignment-potential}
V_{\rm align} =
- \sum^{N-1}_{i=1} \Lambda^4_i 
\cos\left( 
\frac{\phi_i}{f_i} + n_i \frac{\phi_{i+1}}{f_{i+1}} \right),
\eeq
for  $\Lambda_i\gg \Lambda_{\rm QCD}$, where $n_i$ ($i\leq N-1$) are integers, and we define $n_N=1$
for notational convenience.
The above potential provides masses to $N-1$ axions, and there remains one flat direction,
\beq
\label{AlignedAxion}
a = \frac{1}{f_a} \sum^N_{i=1} (-1)^{i-1} 
\left(\prod^{N}_{j=i} n_j \right) f_i \phi_i, 
\eeq
which we identify with the QCD axion.
The effective axion decay constant, $f_a$, is given by
\beq
f_a = \sqrt{\sum^N_{i=1}
\left(\prod^{N}_{j=i} n^2_j \right) f^2_i},
\eeq
and thus it can be enhanced depending on the values of $n_i$ and $N$. 
For later use, let us show how the QCD axion appears in each axion $\phi_i$:
\beq
\frac{\phi_i}{f_i} = (-1)^{i-1} \left(\prod_{j=i}^{N} n_j\right) \frac{a}{f_a} + \cdots,
\eeq
where the dots represent massive modes. 
Assuming for simplicity that the axions have
\beq
f_i = f_N =f  \quad{\rm and}\quad 
|n_i | = n >0 \quad(i=1,2,...,N-1)
\eeq
one finds that the QCD axion $a $ comes mostly from $\phi_1$ and 
$f_a$ is  exponentially enhanced,
\beq
f_a = \sqrt{\frac{n^{2N}-1}{n^2-1}}\, f  \sim e^{N\ln n} f.
\eeq 
For instance, $f_a$ is enhanced by a factor of $10^{6 - 9}$
for $n=3$ and $N= 14- 20$.

To see how to obtain the alignment potential, we consider $N$ complex scalars 
developing a vacuum expectation value (VEV) 
\beq
\Phi _i = \left( \rho_i + \frac{f_i}{\sqrt2} \right)
e^{i\phi_i/f_i},
\eeq
with $f_i=\sqrt2 \langle|\Phi_i|\rangle$, and $\rho_i$ denoting the saxion.
Here the $\Phi_i$'s are stabilized by the potential preserving the global U$(1)$ symmetry 
associated with each complex scalar, for instance, dominantly by 
\beq
V = \sum^N_{i=1}
\left( -m^2_i |\Phi_i|^2 + \frac{\lambda_i}{4} |\Phi_i|^4 \right),
\eeq
with $m_i\sim f$ and $\lambda_i\sim 1$.\footnote{
One may introduce terms like $\sum_{ij} \lambda_{ij} |\Phi_i|^2 |\Phi_j|^2$, 
which mix the saxions but without affecting the alignment mechanism. 
The following arguments are valid in the presence of such terms as long as they do not destabilize the potential.
Also, the scalars may have quartic couplings with the Higgs field. If the couplings are bounded below,
their VEVs may be close to the weak scale.
}
Then there appear $N$ massless axions.

One way to provide masses to these $N-1$ axions while enhancing the effective decay constant of
the remaining massless combination is to
%
add renormalizable interactions breaking $N-1$ global U$(1)$ 
symmetries~\cite{Kaplan:2015fuy},
\beq
\Delta V = \sum^{N-1}_{i=1}
\epsilon_i \Phi_i \Phi^3_{i+1}
+ {\rm h.c.},
\eeq
for small $\epsilon_i$ not to modify the saxion potential significantly.
Integrating out the saxions, one is led to the axion potential of the form (\ref{alignment-potential})
with
\beq
n_i = 3 \quad{\rm and}\quad \Lambda_i = \left( \frac{\epsilon_i}{2} f_i f^3_{i+1} \right)^{1/4}.
\eeq
Alternatively, one may introduce hidden quarks charged under hidden gauge symmetries, whose
non-perturbative effects generate the axion potential~\cite{Choi:2014rja}.
Both models possess one unbroken global U$(1)$ symmetry, which corresponds to the U$(1)_{\rm PQ}$ symmetry
and is to be explicitly broken by the QCD instanton effects.

The axion coupling to gluons can be induced radiatively from the loops of heavy PQ quarks which 
are color-charged and obtain masses from the VEV of one of $\Phi_i$.
Note that the QCD axion fraction in $\phi_i$ quickly decreases with $i$ as in (\ref{AlignedAxion}).
Considering that the QCD axion should couple weakly to 
 gluons in order to satisfy the astrophysical constraints,
we add PQ quarks $Q + \bar{Q}$ coupled to  $\Phi_N$:
\beq
\label{phiNQQ}
\Delta {\cal L} = y_q \Phi_N \bar Q Q.
\eeq
Then, after integrating out the heavy saxion, axions and the PQ quarks, 
one gets the effective action of the QCD axion
\beq
 {\cal L}_{\rm eff} = 
\frac{\alpha_s}{8\pi}
\frac{a }{F_a}  G^{\mu\nu} \tilde G_{\mu\nu}
+ \cdots, 
\eeq 
where $G^{\mu\nu}$ is the SU$(3)_c$ field strength, and $F_a$ is defined by
\beq
F_a = \frac{f_a}{N_{\rm DW}},
\eeq
with $N_{\rm DW}$ being the domain-wall number determined by the number of PQ quarks.
The ellipsis includes the couplings of the QCD axion to the other SM gauge bosons, which are
suppressed also by $F_a$.

\section{Quality of the Peccei-Quinn symmetry}
\label{sec:3}

The QCD axion provides a natural solution to the strong CP problem by dynamically canceling 
the $\theta$ parameter in  QCD.
This mechanism works when the global PQ symmetry is explicitly broken by QCD instanton effects
and other explicit breaking effects are highly suppressed.
However quantum gravity is widely believed not to respect global symmetries.
This implies that there generically exist Planck-suppressed higher dimensional operators explicitly breaking the PQ symmetry.
Thus it is important to understand why the quality of the PQ symmetry remains good enough to solve the strong CP problem 
at low energy scales despite such quantum gravity effects.
The aligned QCD axion models can naturally explain the high quality of the PQ symmetry because the original 
axion decay constants  are much smaller than the effective decay constant of the QCD axion.

Including the QCD instanton and possible quantum gravity effects, the scalar potential of the QCD axion at 
temperature $T=0$ can be written
\beq
\label{axion-potential}
V_{\rm QCD} = -m^2_{\rm QCD}  F_a^2 \cos\left(\frac{a }{F_a}\right)
-  m^2_{\cancel{\rm PQ}\,}\mu^2 \cos\left( \frac{a }{\mu} - \alpha \right),
\eeq 
where the first term is generated by QCD instanton effects
\beq
\label{mf}
m_{\rm QCD} \simeq 6\times 10^{-4}{\rm eV} 
\left(\frac{\Lambda_{\rm QCD}}{400{\rm MeV}}\right)^2
\left(\frac{F_a}{10^{10}{\rm GeV}}\right)^{-1},
\eeq
while the second term represents explicit PQ breaking originating from 
quantum gravity effects. Here we take $0 \leq \alpha \leq \pi$ without loss of generality.  
The minimum of the QCD axion potential is deviated from the origin due to the second term,
and the strong CP phase is estimated to be 
\beq
\label{thetabar}
\bar\theta \equiv  \frac{\langle a \rangle}{F_a}
 &\simeq& \frac{m^2_{\cancel{\rm PQ}}  \sin\alpha }
{m^2_{\rm QCD}+ m^2_{\cancel{\rm PQ}} \cos\alpha}
\frac{\mu}{F_a},
\eeq
for $m^2_{\cancel{\rm PQ}}\sin\alpha\ll m^2_{\rm QCD}+ m^2_{\cancel{\rm PQ}} \cos\alpha$,
and then the total axion mass is determined by
\beq
m^2_a \simeq m^2_{\rm QCD} + m^2_{\cancel{\rm PQ}} \cos\alpha.
\eeq 
This is the case where the QCD axion is stabilized mainly by the first term in (\ref{axion-potential}),
and small shift of the minimum is induced by the second term. 
The CP phase $\bar\theta$ should be smaller than about $10^{-10}$ in order not to generate too large neutron 
electric dipole moment. 
In the conventional QCD axion scenarios, $\mu$ is similar to $F_a$ in size as they are both determined by
the VEV of PQ scalars, implying that $m_{\cancel{\rm PQ}}$ should be highly suppressed compared to 
$m_{\rm QCD}$ to satisfy the experimental constraint on the strong CP violation angle.
However, for instance, a dimension-five Planck-suppressed PQ breaking operator gives 
\beq
m_{\cancel{\rm PQ}} \sim 10^6 {\rm GeV}
\left(\frac{\mu}{F_a}\right)^{3/2} 
\left(\frac{F_a}{10^{10}{\rm GeV}}\right)^{3/2},
\eeq
thereby badly spoiling the PQ mechanism for $\mu\sim F_a$ unless $\alpha$ is extremely small. 
It is obvious that the situation significantly changes if 
$\mu \ll F_a$, 
which is indeed the case in the alignment axion models.

On the other hand, if $m_{\rm QCD}$ is not much larger than $m_{\cancel{\rm PQ}}$,
the relation
$m^2_{\cancel{\rm PQ}}\sin\alpha\ll m^2_{\rm QCD}+ m^2_{\cancel{\rm PQ}} \cos\alpha$ may 
not hold for $\alpha \sim 1$. 
In such a case it is the second term in the axion potential (\ref{axion-potential}) that plays an important role 
in stabilizing the QCD axion.
Then, fixed at the global minimum, the QCD axion has 
\beq
\bar\theta \approx \alpha \frac{\mu}{F_a},
\eeq
for  $\mu \ll F_a$, and its mass is roughly given by $m_{\cancel{\rm PQ}}$. 
Note that in this case the axion potential develops many local minima, making
it difficult to implement the PQ mechanism unless the QCD axion is set to be close to the origin 
in the early Universe. 
We will return to this issue in Section~\ref{sec:4}.

\subsection{Planck-suppressed dimension-five operator}
\label{sec:dim5}
Let us examine if the PQ symmetry can be robust against quantum gravity effects 
in the aligned QCD axion model. 
Among Planck-suppressed higher dimensional operators, those of $\Phi_1$ 
give the dominant contributions to the QCD axion mass.
The dimension-five operator of $\Phi_1$,
\beq
\Delta V_5 = \frac{\kappa_5}{5}\frac{\Phi^5_1}{M_p} + {\rm h.c.},
\label{d5}
\eeq
generates additional axion potential, which takes the form of the second term in 
(\ref{axion-potential}) with 
\beq
m^2_{\cancel{\rm PQ}} = \frac{5 |\kappa_5|}{2\sqrt2} \frac{f^3_1}{M_p},
\quad
\mu= \frac{f_1}{5},
\quad{\rm and}\quad
\alpha = {\rm arg}[\kappa_5].
\eeq
Here $M_p\simeq 2.4\times 10^{18}$~GeV is the reduced Planck mass.
It is easy to see
\beq
m_{\cancel{\rm PQ}} \simeq 
0.03 {\rm MeV} 
\sqrt{|\kappa_5|}
\left(\frac{f_1}{10^3{\rm GeV}}\right)^{3/2},
\eeq
and so the induced mass is much larger than $m_{\rm QCD}$ unless one takes very tiny $\kappa_5$ or 
small $f_1$.
For $m_{\cancel{\rm PQ}}\gg m_{\rm QCD}$, the strong CP violation angle is simply determined by 
\beq 
\bar\theta \approx 
2\times 10^{-10}
\left(\frac{\alpha}{0.1}\right)
\left(\frac{F_a/f_1}{10^8}\right)^{-1},
\eeq 
showing that the experimental bound on $\bar\theta$  can be avoided if 
the term (\ref{d5}) is the only PQ violating term. 
Note that there are many (local) minima for the QCD axion, and the initial position must be
close to $\bar \theta = 0$ to satisfy the neutron EDM bound. This requires a fine-tuning of order $10^{-10}$.
We shall discuss in the next section how the domain wall dynamics can alleviate the tuning of the initial
condition.

In general there may be many other dimension-five operators. 
For instance, the following operator of $\Phi_N$ is also dangerous:
\beq
\Delta V_5' = \frac{\kappa_5'}{5}\frac{\Phi^5_N}{M_p} + {\rm h.c.},
\label{d5N}
\eeq
which takes the form of the second term in 
(\ref{axion-potential}) with 
\beq
m^2_{\cancel{\rm PQ}} = \frac{5 |\kappa_5'|}{2\sqrt2} \frac{f^5_N}{M_p f_a^2},
\quad
\mu= \frac{f_a}{5},
\quad{\rm and}\quad
\alpha = {\rm arg}[\kappa_5'].
\eeq
Then, its contribution to the QCD axion mass reads
\beq
m_{\cancel{\rm PQ}} = 3 \times 10^{-3}{\rm \,eV}\sqrt{|\kappa_5'|}
\left(\frac{f_N}{10^3{\rm GeV}}\right)^{5/2}
\lrfp{f_a}{\GEV{10}}{-1},
\eeq
which is comparable to $m_{\rm QCD}$ for $|\kappa^\prime_5| \sim 1$ and $f_N$ around TeV, 
and its contribution to $\bar \theta$ is not suppressed 
by $f_N/f_a$ differently from the case of $\Delta V_5$.
 
To summarize, among various dimension-five PQ breaking terms, $\Delta V_5$ generates many local minima which
requires a significant fine-tuning of the initial position (unless the domain-wall dynamics is considered; see next section),
and $\Delta V_5'$ gives a too large contribution to the $\theta$ parameter. 
Therefore, the Planck-suppressed dimension-five operators are
dangerous and generically spoil the PQ mechanism in the aligned scenario. 

\subsection{Planck-suppressed dimension-six operator}
 
It is possible to forbid Planck-suppressed operators of odd dimensions by imposing an extra discrete symmetry,
which is presumed to be a remnant of some gauge symmetry in high energy theory.
Here we consider $Z_2$ parity under which all the PQ scalars developing nonzero VEV are odd
\beq
Z_2 \,:\quad
\Phi_i \to -\Phi_i,
\eeq
which is a $Z_2$ subgroup of the U$(1)_{\rm PQ}$ symmetry.
Then one of the most dangerous Planck-suppressed operators is the dimension-six operator of $\Phi_1$:
\beq
\Delta V_6 = \frac{\kappa_6}{6}\frac{\Phi^6_1}{M^2_p} + {\rm h.c.},
\eeq
which generates axion potential of the form of the second term in (\ref{axion-potential}) with 
\beq
m^2_{\cancel{\rm PQ}} = \frac{3 |\kappa_6|}{2} \frac{f^4_1}{M^2_p},
\quad
\mu= \frac{f_1}{6},
\quad{\rm and}\quad
\alpha =\tan^{-1} \left(\frac{{\rm Im}\kappa_6}{{\rm Re}\kappa_6}\right).
\eeq
One thus finds  the strong CP phase to be
\beq
\bar\theta \simeq 1.7\times 10^{-10}
 \frac{m^2_{\cancel{\rm PQ}}  \cos\alpha }{m^2_{\rm QCD} 
+ m^2_{\cancel{\rm PQ}} \cos\alpha}
\left(\frac{\tan\alpha}{0.1}\right)
\left(\frac{F_a/f_1}{10^8}\right)^{-1},
\eeq
where the contribution from $\Delta V_6$ to the axion mass is given by
 \beq
 m_{\cancel{\rm PQ}} \simeq 
 0.5 \times 10^{-3} {\rm eV}
\sqrt{|\kappa_6|}
 \left(\frac{f_1}{10^3{\rm GeV}}\right)^2,
\eeq
which is comparable to or smaller than $m_{\rm QCD}$ for $f_1$ less than TeV.
As can be seen from the relation (\ref{thetabar}), the strong CP violation angle is 
suppressed for $F_a\gg f_1$, and further if $m_{\cancel{\rm PQ}}$ is smaller than
$m_{\rm QCD}$. 
Also, the contributions of terms like $\Phi_N^6/M_p^2$ to the QCD axion mass
are negligibly small.
Note that the model with exact $Z_2$ parity would suffer from the domain-wall problem
if the PQ symmetry breaking occurs after inflation. 
If the $Z_2$ parity is broken by a small amount,  the domain walls will annihilate.

\subsection{Supersymmetric models}
 
Let us briefly discuss explicit PQ breaking effects in the aligned QCD axion within the supersymmetric 
framework.  
To implement the alignment mechanism, we introduce $N$ pairs of the SM-singlet chiral superfields 
$\Phi_i+\hat \Phi_i$.
One way to stabilize them is through the superpotential 
\beq
\mbox{\it model A} :\quad
\Delta W = \sum^N_i X_i \left(
\Phi_i \hat \Phi_i - f^2_i
\right),
\eeq
for $f_i \gtrsim m_{\rm SUSY}$, where $m_{\rm SUSY}$ is the soft SUSY breaking scale, and
we have omitted coupling constants of order unity for simplicity. 
Having soft SUSY breaking scalar masses of similar size, 
$\Phi_i$ and $\hat\Phi_i$ are stabilized as
\beq
\langle |\Phi_i| \rangle \sim \langle |\hat \Phi_i| \rangle \sim f_i,  
\eeq
as required by the $F$-flat condition $|\Phi_i\hat \Phi_i-f^2_i|\sim m^2_{\rm SUSY}$.  
Another way is to consider   
\beq
\mbox{\it model B} :\quad
\Delta W = \sum^N_i \Phi_i \hat \Phi^2_i,
\eeq
with non-tachyonic and tachyonic soft SUSY breaking scalar masses 
for $\Phi_i$ and $\hat \Phi_i$ respectively, and scalar trilinear terms.
Here we have omitted Yukawa couplings.
Then the competition between supersymmetric and SUSY breaking effects leads to    
\beq
\langle |\Phi_i| \rangle \sim \langle |\hat \Phi_i| \rangle \sim m_{\rm SUSY}.   
\eeq 
Note that both models possess  $N$ global U$(1)$ symmetries, and thus there appear $N$ massless 
axions. 
  
Now we add aligning potential terms that give masses to $N-1$ axions and enhance the effective axion 
decay constant of the remaining massless axion.
As in the non-supersymmetric case, one can consider two cases.
One is to add
\beq
\Delta W_{\rm align} = \epsilon \sum^{N-1}_{i=1}\left(
\Phi_i \hat \Phi^2_{i+1}
+ \hat\Phi_i \Phi^2_{i+1} 
\right),
\eeq
for small $\epsilon$.
Then the alignment is achieved with $n_i=2$~\cite{Kaplan:2015fuy}.
Instead one can consider non-perturbative dynamics to get alignment~\cite{Choi:2014rja}:
\beq
\Delta W_{\rm align} =
\sum^{N-1}_{i=1} \left(
\Phi_i \Psi_i \Psi^c_i
+ \sum^{n_i}_{\alpha=1} \Phi_{i+1} \hat\Psi_{i\alpha} \hat\Psi^c_{i\alpha} \right),
\eeq
omitting Yukawa couplings for simplicity. 
Here the hidden quarks $\Psi_i$ and $\hat \Psi_{i\alpha}$ belong to the fundamental representation of hidden
SU$(k_i)$ gauge group which confines at $\Lambda_i$, while $\Psi^c_i$ and $\hat \Psi^c_{i\alpha}$ belong 
to the anti-fundamental representation. 
 
How large are the explicit PQ breaking effects in the supersymmetric models?
As discussed already, the most dangerous Planck-suppressed operators are those of $\Phi_1$ and 
$\hat \Phi_1$ unless one assumes large hierarchy among the original axion decay constants.
Let us first see the effect of Planck-suppressed dimension-four superpotential term:
\beq
\Delta W_{\cancel{\rm PQ}}  = \frac{\xi}{4} \frac{\Phi^4_1}{M_p}.
\eeq
The additional potential of the QCD axion from the above superpotential takes the form (\ref{axion-potential}) 
with 
\beq
m^2_{\cancel{\rm PQ}} = 2|\xi A_\xi| \frac{f^2_1}{M_p},
\quad
\mu= \frac{f_1}{4},
\quad{\rm and}\quad
\alpha =\tan^{-1} \left(\frac{{\rm Im}(\xi A_\xi)}{{\rm Re}(\xi A_\xi)}\right).
\eeq
where $A_\xi\sim m_{\rm SUSY}$ is the soft supersymmetry breaking $A$-parameter associated with 
the $\xi$-term. 
One thus finds that, for $f\sim m_{\rm SUSY}$, the situation is similar to the non-supersymmetric
case with dimension-five operators.
On the other hand, Planck-suppressed dimension-four superpotential terms can be suppressed by 
the gravitino mass,
\beq
\xi \sim \frac{m_{3/2}}{M_p},
\eeq
if gauged U$(1)_R$ symmetry is assumed, i.e., we have $K \ni X^{\dag} \Phi_1^4/M_p^3$
while assigning appropriate U$(1)_R$ charges to $\Phi_i$ and $\hat\Phi_i$.
Here $X$ is the SUSY-breaking field.
Alternatively one may impose some discrete symmetry to forbid dimension-four superpotential terms. 
Then the situation becomes similar to the non-supersymmetric case where Planck-suppressed dimension-five operators are absent due to the $Z_2$ parity.

\section{Cosmology}
\label{sec:4}
Now we study the QCD axion dynamics in the early Universe in the presence of 
PQ breaking terms which
induce extra modulations in the axion potential.  
Before going into details, let us briefly summarize the axion dynamics in the presence of such extra PQ breaking terms.
Most importantly, the QCD axion has a non-zero mass even before the QCD phase transition, and therefore, 
it may start to oscillate earlier than usual. If the PQ breaking terms are sufficiently large, the QCD axion is
trapped in one of the local minima for a long time, 
and the cosmological axion abundance is modified.
In extreme cases, the QCD axion remains trapped in a local minimum which is stable 
during a cosmological time scale, 
and  the PQ mechanism no longer solves the strong CP problem. 
In some case, however, if the QCD axion has sufficiently large quantum fluctuations, or if PQ symmetry is spontaneously 
broken after inflation, 
the minimum closest to $\bar \theta = 0$ satisfying the experimental bound (\ref{theta}) is realized somewhere in the Hubble horizon.
Then, domain walls separating the local minima annihilate with each other, and the almost CP conserving minimum is realized in the
 entire space. This solution to the strong CP problem is different from the ordinary PQ mechanism  in that  quantum fluctuations and domain-wall 
dynamics play an essential role. 

For simplicity, let us consider the axion potential (\ref{axion-potential}) taking account of the temperature dependence of the QCD instanton effects~\cite{Wantz:2009it}
\beq
m_{\rm QCD}^2(T)\;\simeq\;\left\{
\bear{lc}
\ds{c_T~
\frac{\Lambda_{\rm QCD}^4}{F_a^2} \left(\frac{T}{\Lambda_{\rm QCD}}\right)^{-\ell}}
&~~{\rm for~~} T>0.26 \Lambda_{\rm QCD} \\
\ds{c_0~\frac{\Lambda_{\rm QCD}^4}{F_a^2}} 
&~~{\rm for~~}  T<0.26\Lambda_{\rm QCD}
\eear
\right.,
\eeq
with $c_T \simeq 1.68 \times 10^{-7}$, $c_0 \simeq 1.46 \times 10^{-3}$, $\Lambda_{\rm QCD} = 400$\,MeV 
and $\ell=6.68$.
The QCD axion mass $m_{\rm QCD}$ in Eq.~(\ref{mf}) is equal to $m_{\rm QCD}(T=0)$.
So, the potential is given by
\beq
V_{\rm QCD} = -m^2_{\rm QCD}(T)  F_a^2 \cos\left(\frac{a }{F_a}\right)
-  m^2_{\cancel{\rm PQ}\,}\mu^2 \cos\left( \frac{a }{\mu} - \alpha \right).
\eeq 
In general, the axion potential receives corrections from various Planck-suppressed PQ breaking terms, 
in which case the axion potential will be more complicated. Our analysis
however captures the essential features of the axion dynamics, and it can be straightforwardly applied to a more general case if one of 
the PQ breaking operators dominates over the others.

First let us briefly review the case without the extra PQ breaking term, i.e., $m^2_{\cancel{\rm PQ}\,} = 0$. In this case
 the QCD axion starts to oscillate when the mass becomes comparable to the Hubble parameter, 
$m_{\rm QCD}(T_{\rm osc}^{(0)}) \simeq 3 H(T_{\rm osc}^{(0)})$, where the temperature
at the commencement of oscillations is given by
\beq
T_{\rm osc}^{(0)} & =& \left(c_T \frac{10}{\pi^2 g_*} \frac{M_p^2}{F_a^2} \right)^\frac{1}{\ell+4} 
\Lambda_{\rm QCD} \non \\ 
&\simeq& 2.3 {\rm \,GeV} \lrfp{g_*(T_{\rm osc}^{(0)})}{80}{-0.094}\lrfp{F_a}{10^{10}\,{\rm GeV}}{-0.19}.
\eeq
Here $g_*(T)$ counts the relativistic degrees of freedom in plasma with temperature $T$. 
The axion mass at $T=T_{\rm osc}^{(0)}$  is given by
\beq
m_{\rm QCD}(T_{\rm osc}^{(0)}) &\simeq& 2 \times 10^{-8}{\rm\,eV} \lrfp{g_*(T_{\rm osc}^{(0)})}{80}{0.31} \lrfp{F_a}{10^{10}{\rm GeV}}{-0.37}. 
\eeq
The coherent oscillations of the axion contribute to DM, and
its abundance is given by
\beq
\Omega_a h^2 \simeq 0.2\, \theta_i^2 \lrfp{F_a}{\GEV{12}}{1.18},
\label{Omegaa}
\eeq
where $\theta_i$ is the initial misalignment angle. 

Now let us study how the axion abundance is modified in the presence of the PQ breaking term.
The axion cosmology in the presence of the extra PQ breaking term can be broadly classified as follows. 
If  $m_{\cancel{\rm PQ}}^2 \lesssim m_{\rm QCD}^2(T_{\rm osc}^{(0)})$, the QCD axion dynamics is not significantly
modified. In particular, the axion abundance is still approximately given by Eq.~(\ref{Omegaa}).
On the other hand, if $m_{\cancel{\rm PQ}}^2 \gtrsim m_{\rm QCD}^2(T_{\rm osc}^{(0)})$, the QCD axion starts to 
oscillate earlier. The fate of the QCD axion depends on whether the extra PQ breaking term generates multiple
local minima at $T = 0$. 
Roughly speaking, if  $m_{\cancel{\rm PQ}}^2 \lesssim m_{\rm QCD}^2$, there is a unique potential minimum
(up to the domain wall number $N_{\rm DW}$). 
If $m_{\cancel{\rm PQ}}^2 \gtrsim m_{\rm QCD}^2$, on the other hand,
there are many local minima at $T=0$. In the following we consider these cases in turn.

\subsection{Case of $m_{\cancel{\rm PQ}}^2 \lesssim m_{\rm QCD}^2(T_{\rm osc}^{(0)})$}

If the size of the Planck-suppressed operators is suppressed somehow by e.g. additional
discrete symmetries, the axion potential is dominated by the one from the QCD
instanton effects when the axion starts to oscillate.   
The PQ breaking terms may induce a non-zero strong CP phase, but the axion dynamics is essentially same
as in the conventional case. In particular, the axion abundance as well as its isocurvature perturbations 
(if any) are not modified significantly by the PQ breaking terms.

\subsection{Case of $m_{\rm QCD}^2(T_{\rm osc}^{(0)}) \lesssim m_{\cancel{\rm PQ}}^2 \lesssim m_{\rm QCD}^2 $}

If  $m_{\cancel{\rm PQ}}^2 \gtrsim m_{\rm QCD}^2(T_{\rm osc}^{(0)})$, 
the axion feels its non-zero mass and starts to oscillate even before the QCD phase transition. This takes place 
when the mass becomes comparable to the Hubble parameter, $m_{\cancel{\rm PQ}} \simeq 3 H(T_{\rm trap})$, with
\beq
T_{\rm trap} \simeq 170{\rm\,GeV} \lrfp{g_*(T_{\rm trap})}{80}{-\frac{1}{4}} \lrfp{m_{\cancel{\rm PQ}}}{10^{-4}{\rm\,eV}}{\frac{1}{2}},
\eeq
and then the axion is trapped in one of the minima, $a = a_i$, satisfying
\beq
\sin\left(\frac{a_i}{\mu} - \alpha \right) = 0.
\eeq
The abundance of the axion coherent oscillations about $a=a_i$ is given by
\beq
\Omega_{a,{\rm trap}} h^2 \simeq 8.6 \times 10^{-21}\, 
\theta_{\rm trap}^2 \lrfp{g_*(T_{\rm trap})}{80}{-\frac{1}{4}} \lrfp{\mu}{10^{3}{\rm\,GeV}}{2}
\lrfp{m_{\cancel{\rm PQ}}}{10^{-4}{\rm\,eV}}{\frac{1}{2}},
\label{Omegaa1}
\eeq
where $\theta_{\rm trap} = a_i/\mu$ is the initial misalignment measured from $a=a_i$
and we have assumed $m_{\rm QCD} \sim m_{\cancel{\rm PQ}}$ 
at low temperature, but the result will not drastically 
change even for $m_{\rm QCD} > m_{\cancel{\rm PQ}}$.
The abundance
is much smaller than the contribution (\ref{Omegaa}) because it starts to oscillate earlier and
its initial oscillation amplitude is of order $f (\ll F_a)$, and so, we neglect the initial abundance in the following.

During the QCD phase transition, the axion gradually acquires a mass from the QCD instanton effects.
At $T=0$, the axion potential has a unique potential minimum 
(up to the domain wall number $N_{\rm DW}$). This is because the curvature of the potential is dominated by the QCD instanton effects,
and the modulations are subdominant, $m_{\cancel{\rm PQ}}^2 \lesssim m_{\rm QCD}^2$.
The local minimum at $a = a_i$ is destabilized when  the plasma temperature drops down to
\beq
T_{\rm ds} &= & \left(c_T \sin \theta_i \frac{\Lambda_{\rm QCD}^4}{F_a \mu m_{\cancel{\rm PQ}}^2} \right)^\frac{1}{\ell} \Lambda_{\rm QCD}\non\\
&\simeq& 2~ (\sin\theta_i)^{0.15}  \lrfp{F_a}{10^{10}{\rm\,GeV}}{-0.15} \lrfp{m_{\cancel{\rm PQ}}}{10^{-4}{\rm\,eV}}{-0.3} \,{\rm GeV},
\eeq
where $\theta_i \equiv a_i/F_a$.

\begin{figure}[t]
\begin{center}
\begin{minipage}{16.4cm}
\centerline{
{\hspace*{0cm}\epsfig{figure=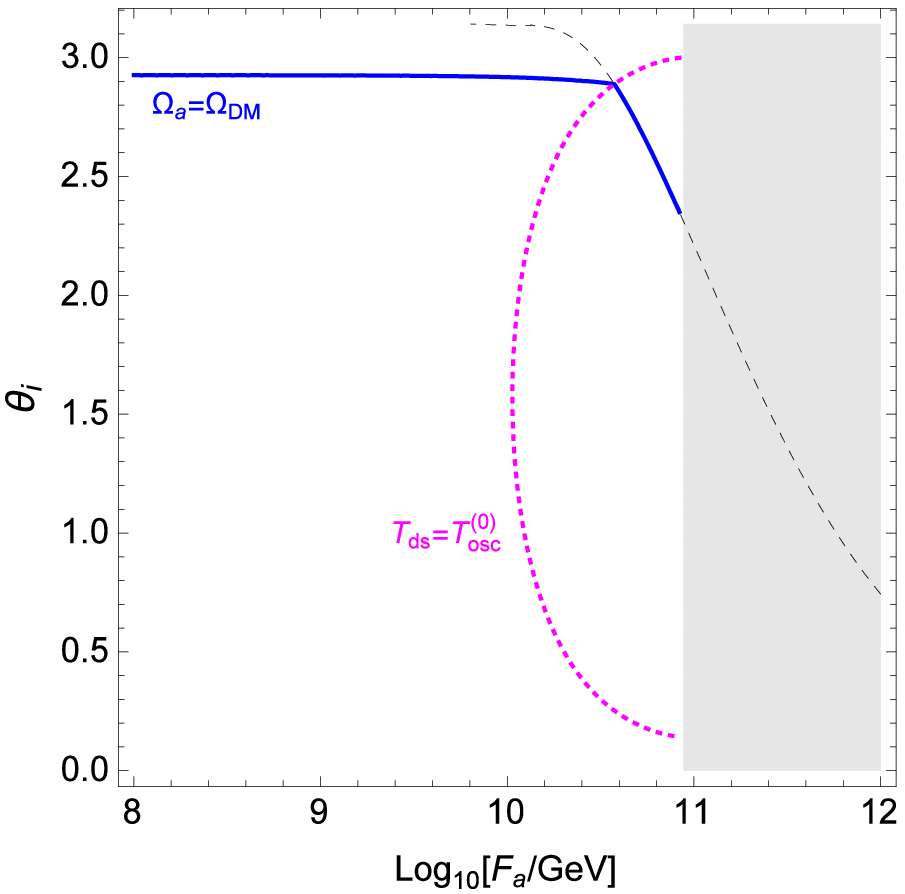,angle=0,width=7.8cm}}
{\hspace*{.4cm}\epsfig{figure=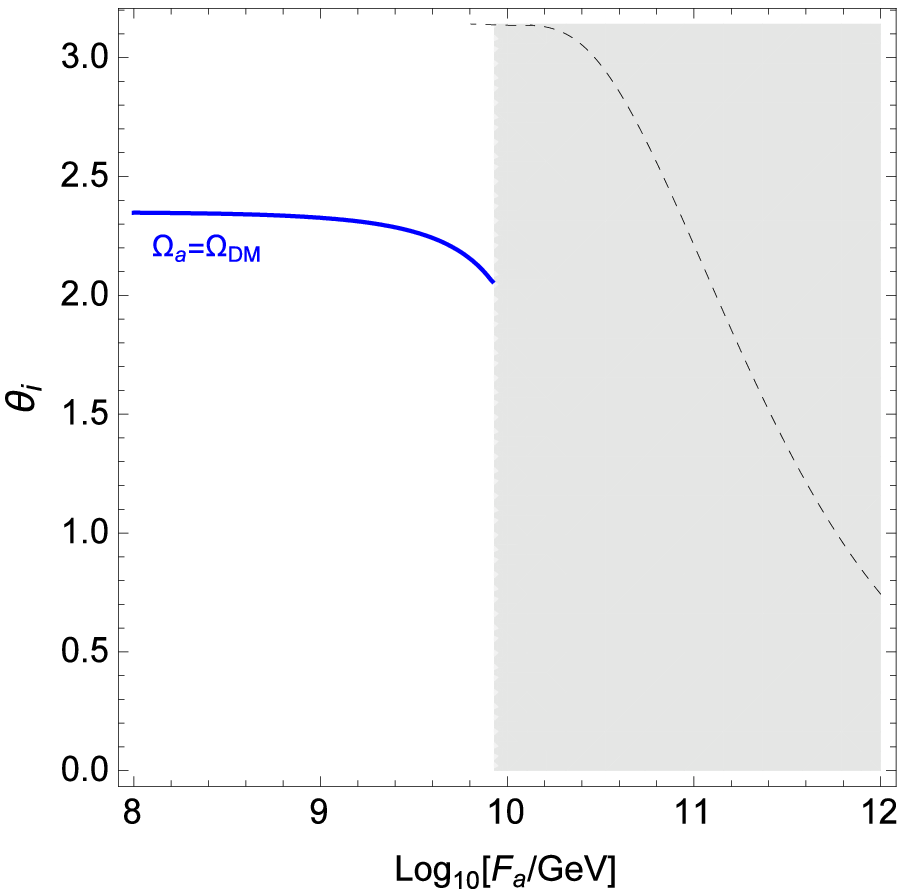,angle=0,width=7.8cm}}
}
\caption{
Axion dark matter density in the aligned QCD axion model in the presence of Planck-suppressed
PQ breaking operators.
Here we have fixed the strong CP phase to be $|\bar\theta|=10^{-10}$,  and taken 
$f=5\mu=1$~TeV and $\sin\alpha=0.1$ ($0.01$) in the left (right) panel.
The axions produced by the misalignment mechanism have $\Omega_a h^2 \simeq 0.1$ 
along the solid blue line, and  $T^{(0)}_{\rm osc}$ is larger than $T_{\rm ds}$ on the left side
of the magenta dotted curve.
For comparison, the contour $\Omega_a h^2 \simeq 0.1$ in the conventional scenario
is shown by the dashed black line. 
In the shaded region, where the QCD axion mass is dominated by the contribution from the Planck-suppressed
PQ breaking operators, the PQ mechanism does not work because there are multiple
local minima.
}
\label{fig:axion1}
\end{minipage}
\end{center}
\end{figure}

If $T_{\rm ds} > T_{\rm osc}^{(0)}$, the axion does not significantly evolve and stays
around $a=a_i$ until the temperature  decreases down to $T_{\rm osc}^{(0)}$. In this case, the axion abundance is given by (\ref{Omegaa}).
If $T_{\rm ds} < T_{\rm osc}^{(0)}$, on the other hand, the axion remains trapped in the local minimum  
even after the temperature becomes lower than $T_{\rm osc}^{(0)}$ at which the axion would start to oscillate
in the absence of the PQ breaking terms.  We can combine the condition $T_{\rm ds} < T_{\rm osc}^{(0)}$ and $\bar \theta \lesssim 10^{-10}$
to derive
\beq
F_a \lesssim 1.6 \times 10^9 (\sin\theta_i \sin\alpha)^{-0.8} \lrfp{g_*(T_{\rm ds})}{80}{-\frac{1}{2}}  \lrfp{\bar \theta}{10^{-10}}{0.8}  {\rm\,GeV},
\eeq
where we have used Eq.~(\ref{thetabar}) assuming $m_{\cancel{\rm PQ}}^2 \lesssim m_{\rm QCD}^2$.
Therefore, if the product of the two angles $\theta_i$ and $\alpha$ is small, the temporal axion trapping takes place
for the axion decay constant in the classical axion window without contradicting the neutron EDM bound.
 The axion abundance in this case is given by
\beq
\Omega_a h^2 &\simeq& 1\times 10^{-3} \lrfp{g_*(T_{\rm ds})}{80}{-1} \theta_i^{1.05}
 \lrfp{m_{\cancel{\rm PQ}}}{10^{-4}{\rm\,eV}}{1.9}  \lrfp{\mu}{1 {\rm\,TeV}}{0.95}  \lrfp{F_a}{10^{10}{\rm\,GeV}}{0.95},\\
 &\simeq&4 \times 10^{-5}\lrfp{g_*(T_{\rm ds})}{80}{-1} \theta_i^{1.05} (\sin \alpha)^{-0.95} \lrfp{\bar{\theta}}{10^{-10}}{0.95},
 \label{omegaacaseb}
\eeq
where we have approximated $\theta_i < 1$ 
and neglected anharmonic effects. 
Interestingly, when expressed in terms of the strong CP phase in the present vacuum, 
the axion abundance becomes independent of the axion decay constant. For instance, if $\alpha$ is sufficiently small,
the axion can be the dominant component of DM even for $F_a \sim 10^9$~GeV. 
The axion trapping can enhance the final axion abundance.

\begin{figure}[t]
\begin{center}
\begin{minipage}{16.4cm}
\centerline{
{\hspace*{0cm}\epsfig{figure=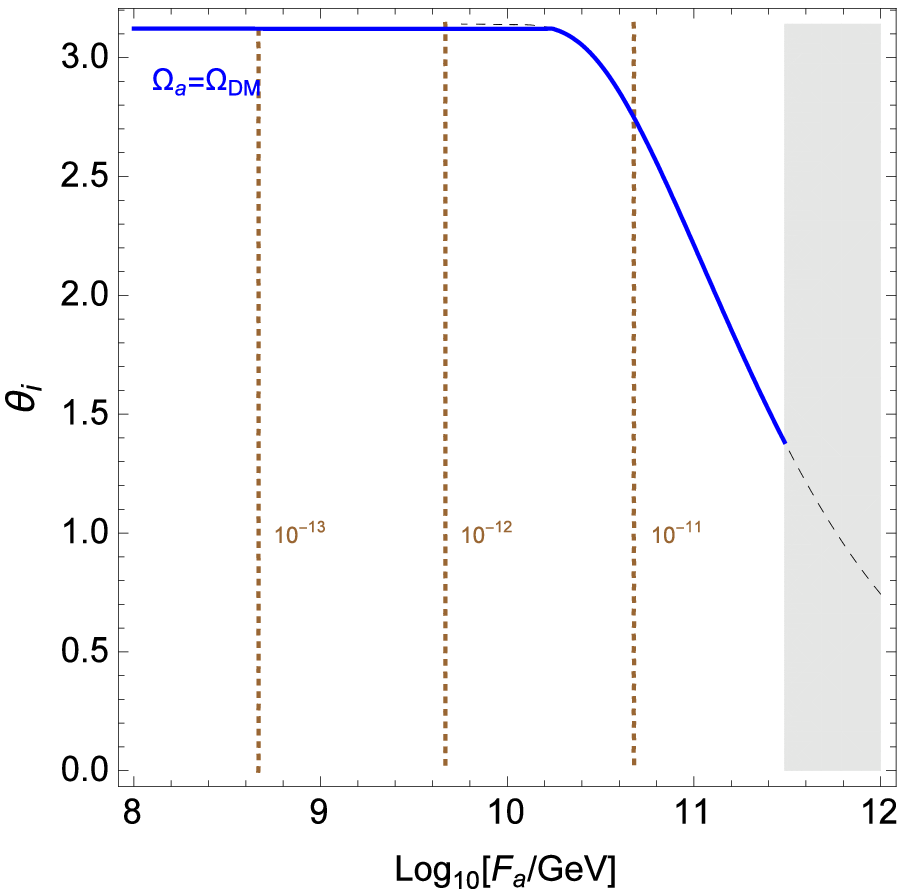,angle=0,width=7.8cm}}
{\hspace*{.4cm}\epsfig{figure=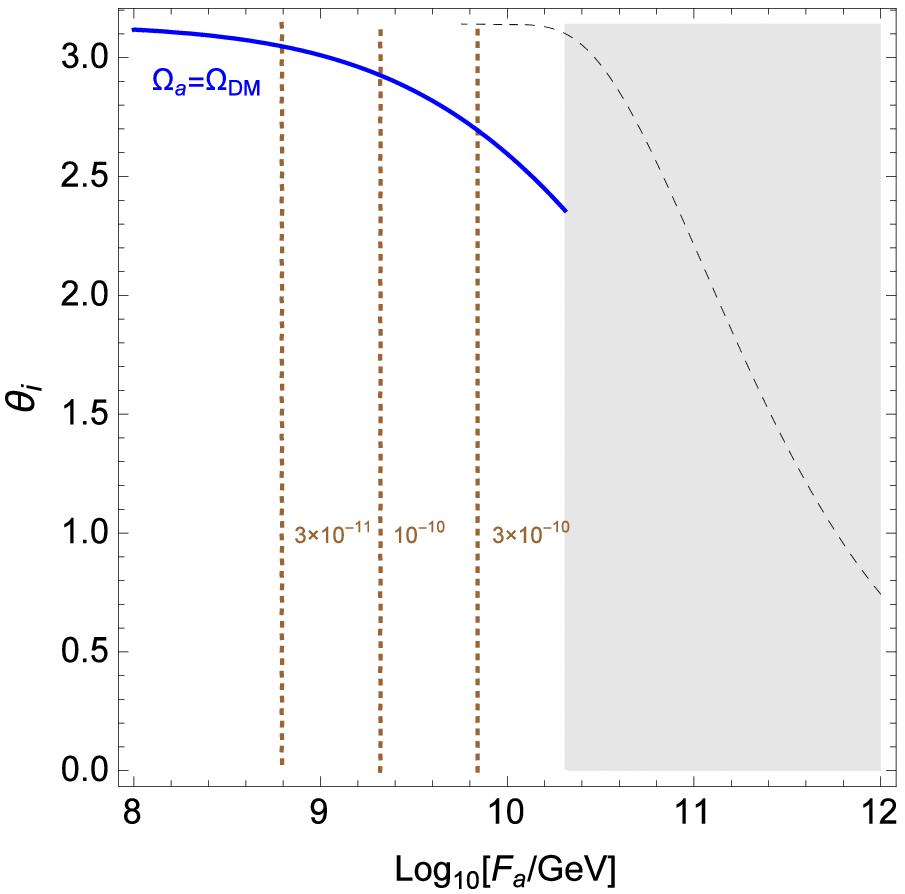,angle=0,width=7.8cm}}
}
\caption{
Axion dark matter density in the aligned QCD axion model in the presence of Planck-suppressed 
PQ breaking operators.
The QCD axion gives $\Omega_a h^2 \simeq 0.1$ along the solid blue line, where 
we have taken $f=5\mu=1$~TeV, $\sin\alpha=0.1$ and $m_{\cancel{\rm PQ}}=2\times 10^{-5}$~eV
($4\times 10^{-4}$~eV) in the left (right) panel.
The brown dotted lines are the contours of $|\bar\theta|$. 
In the shaded region the QCD axion mass is dominated by the contribution from the Planck-suppressed
PQ breaking operators, and the PQ mechanism does not work.
We also show the contour $\Omega_a h^2 \simeq 0.1$ in the conventional scenario
by the dashed blue line. 
}
\label{fig:axion2}
\end{minipage}
\end{center}
\end{figure}

Fig.~\ref{fig:axion1} shows the constant contour of $\Omega_a h^2 = 0.1$ on the $(F_a,\theta_i)$ plane 
in the aligned QCD axion model, where the strong CP phase induced by Planck-suppressed PQ
breaking operators is fixed to be $|\bar\theta|=10^{-10}$.
As noticed above, for fixed $\bar\theta$, the axion abundance becomes insensitive to $F_a$ when the axion 
is trapped in a local minimum even after the temperature drops below $T_{\rm osc}^{(0)}$, i.e.~when
$T_{\rm ds} < T_{\rm osc}^{(0)}$.
On the other hand, Fig.~\ref{fig:axion2} shows the strong CP phase and the axion abundance 
for fixed $m_{\cancel{\rm PQ}}$.
The anharmonic effects have been included in the analysis.

If the PQ symmetry is restored during or after inflation, there is no axion isocurvature perturbation.
If not, the axion acquires quantum fluctuations of order $H_{\rm inf}/2\pi$ during inflation,
where $H_{\rm inf}$ is the Hubble parameter during inflation.
 In the present case, the evolution of  isocurvature perturbation is slightly involved. If the typical size of the quantum fluctuations is smaller than
$2\pi \mu$ and the initial position of the axion is not close to the top of the potential,
the isocurvature perturbations are suppressed because the axion starts to oscillate earlier
and its oscillation amplitude becomes smaller and smaller as the Universe expands until the local minimum
is destabilized by the QCD instanton effects. The energy density associated with the isocurvature perturbations
is  smaller than or comparable to the contribution (\ref{Omegaa1}).
On the other hand, if the initial position of the axion is close to the local maxima, or 
if the typical size of the quantum fluctuations is larger than $2\pi \mu$,\footnote{
If $H_{\rm inf} \gg \mu$, (some of) the scalars $\Phi_i$ may be stabilized at the origin. Even if $H_{\rm inf} \lesssim \mu$,
the quantum fluctuations can be enhanced by the resonant behavior between two axions~\cite{Daido:2015bva}.} the axion is trapped in different minima at different spatial points.
The axion perturbations $\delta a(x)$ become highly non-Gaussian
because the axion takes only discrete values when it is trapped in local minima within the original Gaussian fluctuation. The typical magnitude of the axion 
perturbations at super-horizon scales is retained for the moment 
and still given by $\sim H_{\rm inf}/2\pi$. However, domain walls are formed soon after the axion
gets trapped at different minima. The domain walls will quickly follow the scaling law, which implies that each Hubble horizon contains
one or a few domain walls of the same type. As a result, the isocurvature perturbations at super-horizon scales
are considered to be suppressed by the domain-wall dynamics, even though the axion perturbations at 
subhorizon scales are of order unity.\footnote{
Similar suppression of the isocurvature perturbations was pointed out in Ref.~\cite{Daido:2015gqa} in the context
of spontaneous baryogenesis due to axion domain walls.
} (Here the size of the axion perturbations is measured in units
of $\mu$ until the QCD instanton effects turn on.)
The domain walls annihilate and disappear when those local minima are destabilized by the QCD instanton effects, and their contribution
to the final axion density is considered to be smaller than the coherent oscillations (\ref{omegaacaseb}) owing to $\mu \ll F_a$
unless the initial misalignment angle is very small. Thus, the axion isocurvature perturbations at CMB scales
can be suppressed by the scaling behavior of the domain wall dynamics.

\subsection{Case of $m_{\cancel{\rm PQ}}^2 \gtrsim m_{\rm QCD}^2$}
Now we consider a case in which the PQ symmetry breaking terms give the dominant contributions to
the QCD axion mass even at zero temperature. The strong CP phase at the minimum closest to $\bar \theta = 0$ can satisfy the neutron EDM bound
if the enhancement of the decay constant due to the alignment is sufficiently large, and/or if $\alpha$ is  mildly fine-tuned,
as we have seen in Sec.~\ref{sec:3}. 

Question is how the axion is stabilized at the minimum
with the smallest $\bar \theta$.
There are many ($\sim f_a/\mu$) local minima in the axion potential, and therefore, the
original idea of the PQ mechanism, namely, the dynamical cancellation of the strong CP phase, does not
work. The axion will be simply trapped in the nearest minimum
from the initial position, and it will stay there until the tunneling into the adjacent
minimum takes place. Unless the initial position of the axion is set miraculously close to $\bar\theta = 0$,
 the Universe will be dominated by the axion potential energy and continues to expand exponentially.
The cosmological catastrophe, however, can be avoided if the axion quantum fluctuations are sufficiently large
so that the true minimum close to $\bar \theta = 0$ is realized somewhere in the Hubble horizon.\footnote{
Here we assume that the minimum closest to $\bar \theta = 0$ is the global minimum. This may not be the
case in the presence of other Planck-suppressed operators, as we have seen in Sec.~\ref{sec:dim5}.
} 
This is the case
if the PQ symmetry is restored during or after inflation, or if the quantum fluctuation dominates over the classical
value.  Then, the true minimum close to $\bar \theta = 0$ 
will be realized in the whole Universe
when domain walls annihilate after the QCD instanton effect turns on. In the following we shall study this scenario,
focusing on the domain wall dynamics.

First, let us suppose that the quantum fluctuations dominate over the classical field value so that
$a_i \pm \delta a$ contains  the minimum  closest to $\bar \theta = 0$.
Later we will briefly discuss the case in which the PQ symmetry is spontaneously broken after inflation.
When the axion starts to oscillate,  domain walls are formed, separating various minima inside the axion field fluctuations.
The tension of domain walls, $\sigma$, is given by
\beq
\sigma \simeq 8 m_{\cancel{\rm PQ}} \mu^2.
\eeq
Once formed, domain walls will quickly follow the scaling law. In order to avoid the cosmological domain wall problem,
those domain walls must disappear when the QCD instanton effect turns on and the energy bias between different
minima is induced. The domain walls disappear when its energy density becomes comparable to the energy bias,
and therefore, domain walls separating the minima with the smallest energy bias will be the most long-lived.
Such domain walls connect the minimum closest to the (almost) CP conserving one, $a_0 \simeq \alpha \mu$, and the adjacent one,
$a_1 \simeq a_0 - 2\pi\mu$. 
The strong CP phase at $a=a_0$ is given by
\beq
\bar \theta \simeq \alpha \frac{\mu}{F_a}.
\eeq
In order to solve the strong CP problem without severe fine-tuning of $\alpha$, one needs large hierarchy between $F_a$ and $\mu$,
and we take $\mu = 1$\,TeV and $F_a = 10^{12}$\,GeV and $\alpha = 0.1$ 
as reference values.
The smallest energy bias $\epsilon_{\rm min}$ is then given by
\beq
\epsilon_{\rm min} \simeq 2\pi m_{\rm QCD}^2(T)  \mu^2 
(\pi - \alpha ) .
\eeq
The domain walls annihilate and disappear when their energy density becomes comparable to the energy bias,
and the Hubble parameter at that time is given by
\beq
H_{\rm ann} &\simeq& \frac{\epsilon_{\rm min}}{\sigma} \non \\
& \simeq& \frac{3\pi}{4} \frac{m_{\rm QCD}^2(T_{\rm ann}) }{m_{\cancel{\rm PQ}}  }, 
\eeq
where we have used the fact that the energy density of domain walls in the scaling regime is approximated by $\rho_{\rm DW} \sim \sigma H$ 
and we have fixed $\alpha = 0.1$.
For $m_{\cancel{\rm PQ}} \gtrsim 10$~eV, the domain walls annihilate after the temperature dependence of the QCD axion mass disappears, i.e. $T < 0.26 \Lambda_{\rm QCD}$.
The annihilation temperature in that case is
\beq 
T_{\rm ann} \simeq 100 {\rm\,MeV} \lrfp{g_*(T_{\rm ann})}{10}{-\frac{1}{4}} \lrfp{m_{\cancel{\rm PQ}}}{10{\rm\,eV}}{-\frac{1}{2}}
\lrfp{F_a}{10^{12}\,{\rm GeV}}{-1},
\eeq
and otherwise 
\beq
T_{\rm ann} \simeq 600 {\rm\,MeV} \left[ \lrfp{g_*(T_{\rm ann})}{80}{-\frac{1}{2}} \lrfp{m_{\cancel{\rm PQ}}}{10^{-4}{\rm\,eV}}{-1} \lrfp{F_a}{10^{12}\,{\rm GeV}}{-2}  \right]^{\frac{1}{\ell+2}}.
\eeq 
Therefore,  the domain walls typically annihilate during or soon after the QCD phase transition.
In particular, even if $m_{\cancel{\rm PQ}} \gg m_{\rm QCD}$, the decay can take place well before the domain walls
start to dominate the Universe. 

The axions are copiously produced when the domain walls annihilate, and their abundance is
\beq
\Omega_{a,{\rm DW}} h^2 &\simeq & 2 \times 10^{-11} \lrfp{g_*(T_{\rm ann})}{10}{-\frac{1}{4}}
\lrfp{m_{\cancel{\rm PQ}}}{10{\rm\,eV}}{\frac{3}{2}}
 \lrfp{\mu}{1{\rm TeV}}{2} \lrf{F_a}{10^{12}\,{\rm GeV}}, 
\eeq
for $T_{\rm ann} < 0.26\Lambda_{\rm QCD}$ and otherwise
\beq
\Omega_{a,{\rm DW}} h^2 \simeq 2 \times 10^{-17} \lrfp{g_*(T_{\rm ann})}{80}{-\frac{\ell+1}{2\ell+4}}
\lrfp{m_{\cancel{\rm PQ}}}{10^{-4}{\rm\,eV}}{\frac{\ell+3}{\ell+2}}
 \lrfp{\mu}{1{\rm TeV}}{2} \lrfp{F_a}{10^{12}\,{\rm GeV}}{\frac{2}{\ell+2}},
\eeq
which is much smaller than the observed DM abundance unless $m_{\cancel{\rm PQ}}$ is very heavy.
Note however that we have here focused on the most long-lived domain walls, and there may be
much larger contributions from the other domain walls which have annihilated before. More detailed analysis
of the domain-wall dynamics with different energy bias is necessary, and we will leave it for future work.

So far we have considered the case in which the PQ symmetry is already broken during inflation.
If PQ symmetry is restored, cosmic strings are formed. In fact, the cosmic string formation is rather complicated
in the aligned axion scenario. This can be understood by noting that each cosmic string of $\Phi_i$ has a tension
of order $f_i^2$, but the cosmic string corresponding to the QCD axion should have a tension of order $F_a^2$.
Such cosmic strings in the aligned axion models have interesting cosmological implications, and
we will discuss them in a separate work~\cite{HJKST}.

\section{Discussion and Conclusions}
\label{sec:5}

The high quality of the PQ 
symmetry is a natural outcome in the aligned QCD axion model.
In this framework, the problem can be rephrased as a question: why such an alignment is realized in nature.
It may be an accidental symmetry due to the locality in extra dimensions~\cite{Izawa:2002qk,Kaplan:2015fuy}.
Alternatively, it may be realized by requiring the axion DM~\cite{Carpenter:2009zs}. 
Indeed, the longevity of DM is a puzzle, and the QCD axion is a plausible candidate if the decay
constant is sufficiently large. If there are many gauge singlet scalars at the weak scale, they may
conspire to generate one very flat direction to generate the right amount of DM. In non-SUSY framework,
there are two natural scales for those scalars. One is the cut-off scale like the Planck or GUT scale because
their masses are unstable against radiative corrections. On the other hand, once allowing the weak scale 
to be realized by the fine-tuning (or anthropic argument), these singlet scalars may also have masses
and VEVs of order the weak scale. This is the case if the quartic coupling between the SM Higgs field
and the singlets are constrained to be non-zero. In this case, the aligned QCD axion model emerges
from the axion landscape at the weak scale~\cite{Higaki:2014pja,Higaki:2014mwa}.

Lastly we mention cosmological implications of the present scenario in which the axion mass is determined by
the PQ symmetry breaking terms. The axion mass can be heavier than the conventional one,
in which case the QCD axion can be thought of as axion-like particles whose mass and decay constant do 
not satisfy the relation (\ref{mf}). 
In extreme cases the axion may be unstable in a cosmological time scale. Suppose that the axion is coupled
to a hidden U(1)$_H$ gauge symmetry. A priori there is no reason to expect that the alignment takes place for the coupling to the hidden
photon, and its interaction may be written as
\beq
{\cal L} = \frac{\alpha_H}{8\pi} \frac{a}{f} F^{(H)}_{\mu \nu}\tilde{F}^{(H)\mu \nu},
\eeq
where $\alpha_H$ is the fine-structure constant for the hidden U(1)$_H$ gauge interaction. The decay rate
of the axion into two hidden photons is
\beq
\tau_{a \to \gamma' \gamma'} \simeq 10^{19}{\rm\,sec} \lrfp{\alpha_H}{1/137}{-2} \lrfp{m_a}{1{\rm\,eV}}{-3}\lrfp{f}{10 \,{\rm TeV}}{2}.
\eeq
This is intriguingly close to the present age of the Universe.  
Such decaying dark matter may improve the tension on $\sigma_8$~\cite{Enqvist:2015ara},
if it constitutes a significant fraction of dark matter.
Similarly,
the axion is coupled to ordinary photons, but we assume that the coupling to photons is suppressed by $F_a$ to avoid
various experimental and astrophysical constraints on the axion. This can be realized if
the coupling to photons is induced only by the interaction like Eq.~(\ref{phiNQQ}). In the absence of the decay channel
into hidden photons, the axion may mainly decay into photons producing a line signal in the X-ray spectrum.
If its mass is about $7$~keV and $F_a \sim 10^{15}$~GeV, such axion may account for the $3.5$~keV X-ray line
signal~\cite{Higaki:2014zua,Jaeckel:2014qea,Lee:2014xua}.

In this paper we have studied in detail the quality of  PQ symmetry in the aligned QCD axion scenario
and its cosmological implications. We have found that the PQ symmetry is much more robust against Planck-suppressed 
higher dimensional operators compared to the conventional axion model. The axion abundance can be significantly 
modified if the axion is trapped in one of the local minima generated by the extra PQ symmetry breaking terms.
Generally we expect that a non-zero strong CP phase is induced by the Planck-suppressed PQ symmetry breaking terms,
and its contribution to the neutron EDM can be close to the current upper bound.

One important difference from the conventional axion model is that the 
axion decay constant is not directly related to the dynamical scale of each PQ scalar in our scenario. 
In fact, the symmetry breaking scale is much lower than the axion decay constant. This leads to an interesting
and important effects on the symmetry restoration and the subsequent formation of topological defects. 
In the conventional axion model, we expect that the PQ symmetry can be restored if the inflation scale or the reheating
 temperature is higher than the axion decay constant. In our case, one should compare the inflation scale or the reheating temperature
with the typical size of $f$, which is much smaller than $F_a$ in the classical axion window.
Therefore, it is more likely that the PQ symmetry is restored during and/or after inflation compared 
to the conventional scenario. In this case, the axion is considered to be copiously produced by the annihilation of the
string-wall network during the QCD phase transition, constraining the axion decay constant
as $F_a = {\cal O}(10^{10})$~GeV~\cite{Kawasaki:2014sqa}.  We note however that the axionic string has
a complicated structure made of sub-strings and domain walls, and their formation and the subsequent evolution are
quite non-trivial. We will study this issue in separate publication~\cite{HJKST}.

As pointed out in Ref.~\cite{Higaki:2015jag}, our scenario predicts many axions and saxions around the weak scale.  
They are coupled to gluons in order for the QCD axion to solve the strong CP problem.  So, one of them can 
account for the recently found $750$~GeV diphoton excess. The aligned QCD axion therefore naturally connects
the $750$~GeV diphoton excess to the QCD axion, and provides various implications for the axion DM.
Further study is clearly warranted.

%
\section*{Acknowledgments}
F.T. thanks T.~Sekiguchi for discussion on the cosmic string formation in the aligned axion model.
This work is supported by MEXT KAKENHI Grant Numbers 15H05889 (F.T.) and 23104008 (F.T.),
JSPS KAKENHI Grant Numbers 24740135 (F.T.),  26247042(F.T. and T.H.), and 26287039 (F.T.),
World Premier International Research Center Initiative (WPI Initiative), MEXT, Japan (F.T.),
and MEXT-Supported Program for the Strategic Research Foundation at Private Universities,
"Topological Science", Grant Number S1511006 (T.H.),
the Max-Planck-Gesellschaft, the Korea Ministry of Education, Science and Technology,
Gyeongsangbuk-Do and Pohang City for the support of the Independent Junior Research Group at the
Asia Pacific Center for Theoretical Physics (N.K.).
This work is also supported by the National Research Foundation of Korea (NRF) grant funded by the Korea government (MSIP)
(NRF-2015R1D1A3A01019746) (K.S.J).
%



\end{document}